\documentclass{elsart}

\usepackage{graphicx}
\usepackage{amssymb}

\newcommand{\Fi}[1]{Fig.~\ref{#1}}

\newcommand{\gevc}{\mbox{GeV$/c$}}

\newcommand{\rb}[1]{\mbox{\textrm{\scriptsize #1}}}
\newcommand{\rbt}[1]{\mbox{\textrm{\tiny #1}}}

\newcommand{\sqrts}{\ensuremath{\sqrt{s_{_{\rbt{NN}}}}}}

\newcommand{\lam}{\ensuremath{\Lambda}}
\newcommand{\lab}{\ensuremath{\bar{\Lambda}}}

\newcommand{\hmin}{\ensuremath{\textrm{h}^-}}
\newcommand{\pimin}{\ensuremath{\pi^-}}
\newcommand{\piplus}{\ensuremath{\pi^+}}

\newcommand{\kzeros}{\ensuremath{\textrm{K}^{0}_{\rb{s}}}}

\newcommand{\pbar}{\ensuremath{\bar{\textrm{p}}}}

\newcommand{\mypt}{\ensuremath{p_{\rb{t}}}}

\newcommand{\ncoll}{\ensuremath{\langle N_{\rb{coll}} \rangle}}

\newcommand{\raa}{\ensuremath{R_{\rb{AA}}}}
\newcommand{\rcp}{\ensuremath{R_{\rb{CP}}}}
\newcommand{\rpa}{\ensuremath{R_{\rb{pA}}}}

\begin{document}

\begin{frontmatter}



\title{High $p_{\rb{t}}$ Measurements at the CERN SPS}


\author{C. Blume}

\address{Institut f\"{u}r Kernphysik, J.W.~Goethe Universit\"{a}t, Frankfurt am Main, Germany}

\begin{abstract}

The current experimental situation concerning high \mypt\ observables at the
CERN SPS is reviewed. Recent data from the NA45, NA49 and NA57 collaborations
are discussed and compared to earlier measurements by WA98 and NA45 at the same
center-of-mass energies, as well as to measurements at the higher energies by the RHIC
experiments. 
The observables include new p+p, A+A spectra, nuclear modification factors (\raa, \rcp),
two particle azimuthal correlations, and baryon to meson ratios at
moderately high \mypt.
Generally, the interpretation of the SPS data suffers from the lack of
reliable baseline measurements (p+p and p+A). However, the overall picture
that is emerging suggests that already at SPS energies medium effects similar to
those observed at RHIC are present.

\end{abstract}

%\begin{keyword}

%\PACS 
%\end{keyword}
\end{frontmatter}

% main text
\section{Introduction}

At RHIC energies a strong suppression of high \mypt\ hadron production in 
nucleus--nucleus collisions relative to p+p collisions has been observed~\cite{phenixe,stare}, 
which is usually attributed to the energy loss that the parent partons experience
when transversing the color deconfined medium present in these collisions~\cite{jetquench}.
Also, two particle correlations at high \mypt\ indicate a strong effect of 
the produced medium on the hadron emission pattern.
At the lower SPS energies on the other hand, the experimental situation was
rather unclear and even contradictory. While, e.g. the energy density derived
from the measured $E_{\rb{t}}$ distributions, as well as the fact that
the chemical freeze-out occurs very close to the theoretically expected phase
boundary, suggest that a quark gluon plasma is already formed at top SPS energies~\cite{spsqgp}, 
there was so far no indication of any high \mypt\ suppression at these energies. 
In fact, the WA98 data on the nuclear modification
factor \raa\ of neutral pions rather exhibited a Cronin--type of enhancement 
\cite{wa98pi0}. However, a reassessment of the p+p baseline \cite{david} allowed 
the conclusion that the data on pion production in central A+A reactions are 
consistent with the expectation in a jet quenching scenario with gluon densities 
of $dN_{\rb{g}}/dy = 400 - 600$ \cite{glv1}, more in line with the measured hadron
multiplicities.
A first study of two particle azimuthal correlations for $\mypt > 1.2$~\gevc\
by the NA45 experiment \cite{na45a}, showed a structure that could be 
interpreted as a correlation of semi-hard particles on top of a collective flow component.
Similar to what has been seen at RHIC, the away-side peaks broadens with
centrality, which might be attributed to in-medium effects.
Recently, new data on \rcp\ for identified particles have been presented by
the NA49 and NA57 experiment \cite{na49a,na57a}. Additionally, the NA45 
collaboration has done a refined analysis of two particle azimuthal correlations
at high \mypt\ \cite{na45b}. These data help to clarify the current situation
and shall be discussed in the following.

\section{Baseline measurements in p+p and p+A}

\begin{figure}[t]
\begin{center}
\begin{minipage}[b]{68mm}
\begin{center}
\includegraphics[height=75mm]{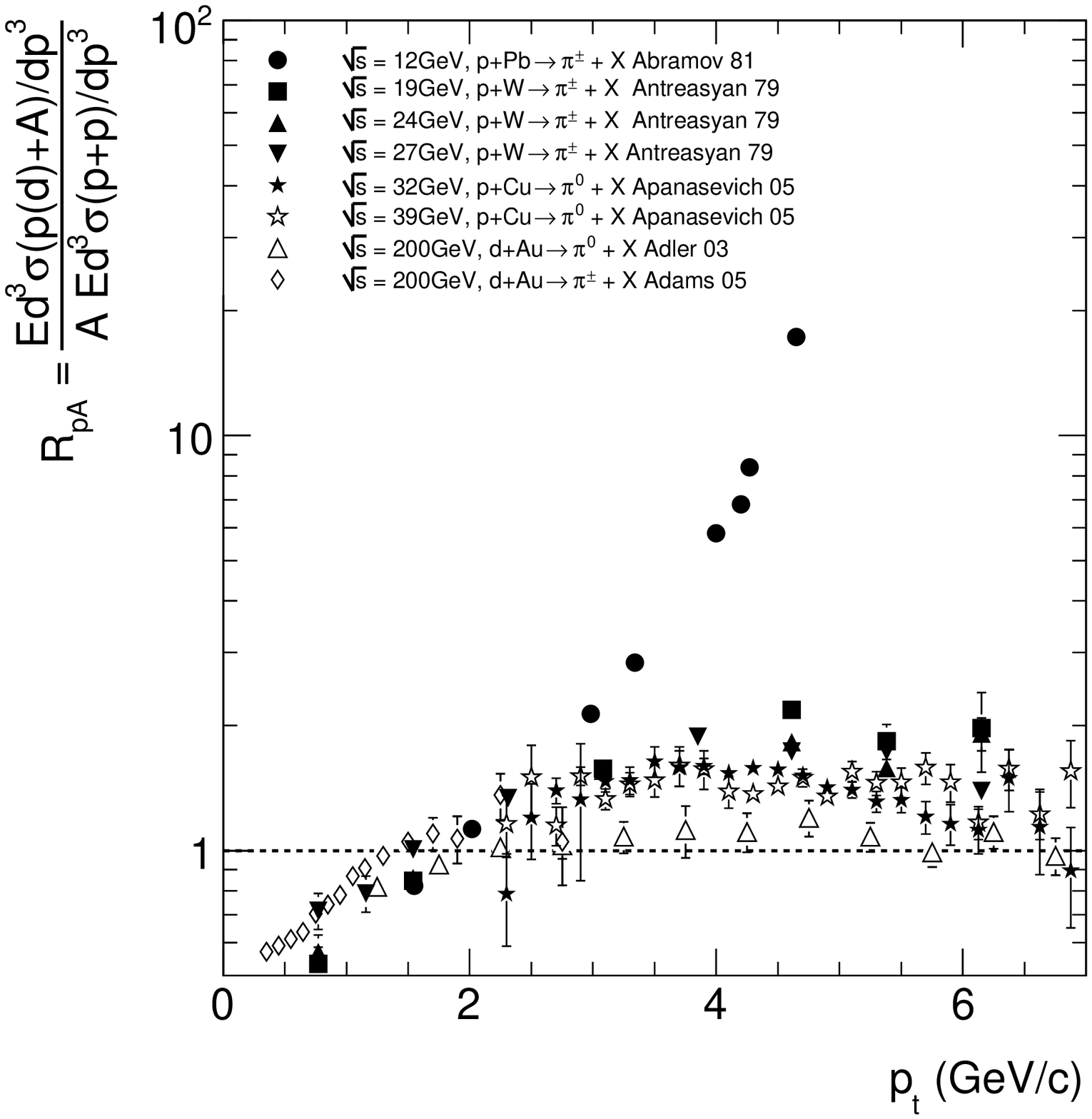}
\end{center}
\end{minipage}
\begin{minipage}[b]{68mm}
\begin{center}
\includegraphics[height=75mm]{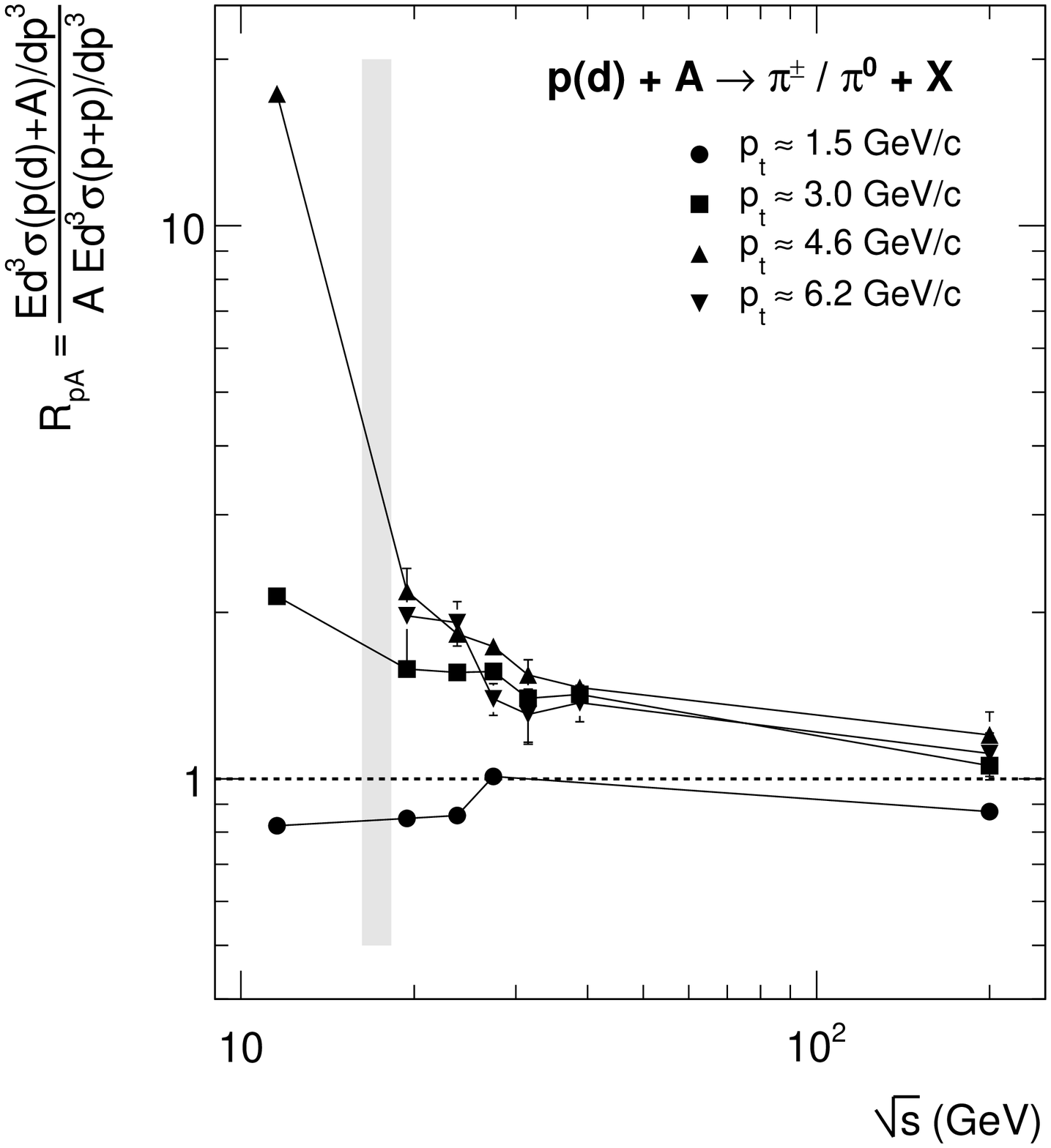}
\end{center}
\end{minipage}
\end{center}
\caption{Left: The nuclear modification factor \rpa\ as a function of \mypt\ for
charged and neutral pions (minimum bias p(d)+A data \cite{abramov,cronin,e706,phenixa,stara}). 
Right: \rpa\ at different fixed momenta as a function of \sqrts. The grey band 
indicates the SPS center-of-mass energy.}
\label{FIrpa}
\end{figure}

\begin{figure}[t]
\begin{center}
\includegraphics[width=100mm]{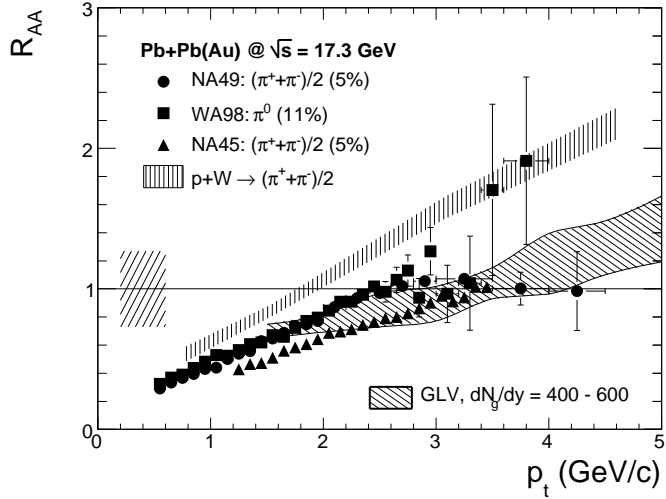}
\end{center}
\caption[]{The nuclear modification factor \raa\ as a function of \mypt\ for neutral
and averaged charged pions \cite{wa98pi0,na45c,na49a}. The hatched box represents 
the typical normalization error 
from the \ncoll\ determination. The lower hatched band is a theory prediction including
energy loss, Cronin effect and nuclear shadowing \cite{glv1}. The upper dashed band shows
\rpa\ as measured in p+W at \sqrts~=~19.4~\gevc\ \cite{cronin}.}
\label{FIraa}
\end{figure}

In order to establish whether any kind of modification in the high \mypt\
region is present in A+A collisions, reference data from p+p and p+A collisions
are of high importance. Unfortunately, there are no p+p(A) data available at 
$\sqrt{s} = 17.3$~GeV that cover the interesting \mypt-region for A+A studies above 2~\gevc.
Recently, the NA49 experiment has published charged pion spectra measured at this
center-of-mass energy \cite{na49b}, but the statistics limits the \mypt-reach
to 2.1~\gevc. Several attempts have been made to replace the missing data by
an interpolation from lower and higher beam energies \cite{wa98pi0,david}. 
In \cite{david} the parametrization suggested by Blattnig et al. \cite{blattnig} 
was used to construct a p+p baseline for neutral pions. This parametrization also 
agrees to the charge averaged pion spectra in the region measured by NA49 at 
$\sqrt{s} = 17.3$~GeV.
However, one should keep in mind that at the center-of-mass energies under
discussion here, the spectral shape in the higher \mypt\ region (i.e. above \mypt\ = 2~\gevc) 
changes drastically with energy since the kinematic limit becomes important here (see e.g. \cite{beier}).
Therefore, any parametrization introduces a large systematic error.
A specific problem in the interpretation of the SPS data arises from the fact that
the typical \mypt-reach is in the region below 4--5~\gevc. This region is
governed by an interplay of the Cronin effect, which causes an enhancement in
p+A collisions relative to p+p, and a potential suppression due to jet
quenching. A proper understanding of the p+A data is therefore indispensable.
This is illustrated by \Fi{FIrpa} that shows the existing data on the
nuclear modification factor \rpa
%, defined as:
%\begin{equation}
%  \rpa = \frac{1}{A} 
%         \frac{E \,d^{3}\sigma(p(d) + A) \,/ \,dp^{3}}{E \,d^{3}\sigma(p + p) \,/\, dp^{3}}
%\end{equation}
at different center-of-mass energies \cite{abramov,cronin,e706,phenixa,stara}.
\rpa\ increases by approximately a factor 2 when going from \sqrts~=~200~GeV down to
\sqrts~=~19.4~GeV at $\mypt \approx 4.6$~\gevc. 
This is not unexpected since due to the steeper spectral shape at lower energies
\rpa\ is more sensitive to \mypt\ broadening effects. 
According to the data from Serpukhov 
at \sqrts~=~11.5~GeV \cite{abramov}, \rpa\ increases towards lower energies 
even more rapidly, making interpolations very difficult. 
%Since any nuclear effect in 
%A+A collisions has to be weighted against the effects already present in p+A, 
%a precise measurement at \sqrts~=~17.3~GeV would be desirable.

\section{Nuclear modification factors in A+A}

\begin{figure}[t]
\begin{center}
\begin{minipage}[b]{68mm}
\begin{center}
\includegraphics[height=62mm]{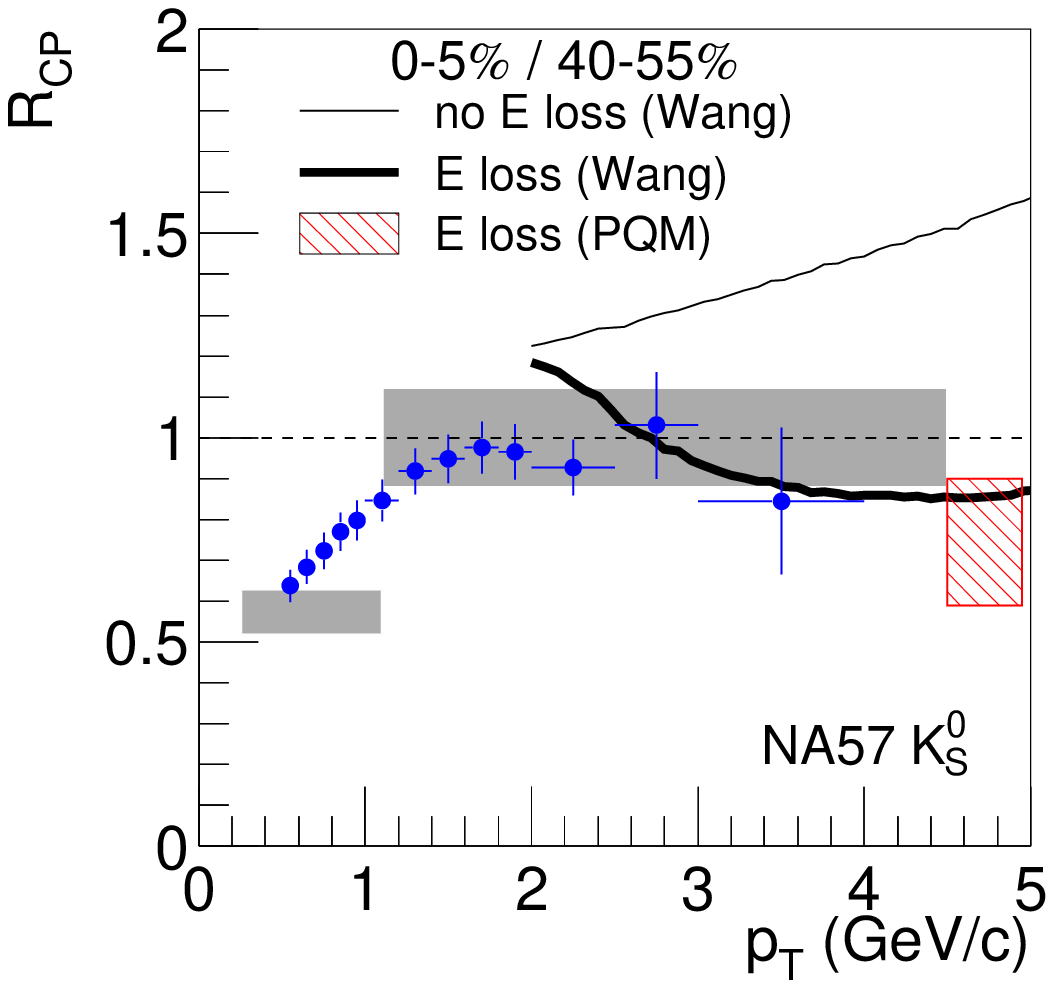}
\vspace{0.1mm}
\end{center}
\end{minipage}
\begin{minipage}[b]{68mm}
\begin{center}
\includegraphics[height=68mm]{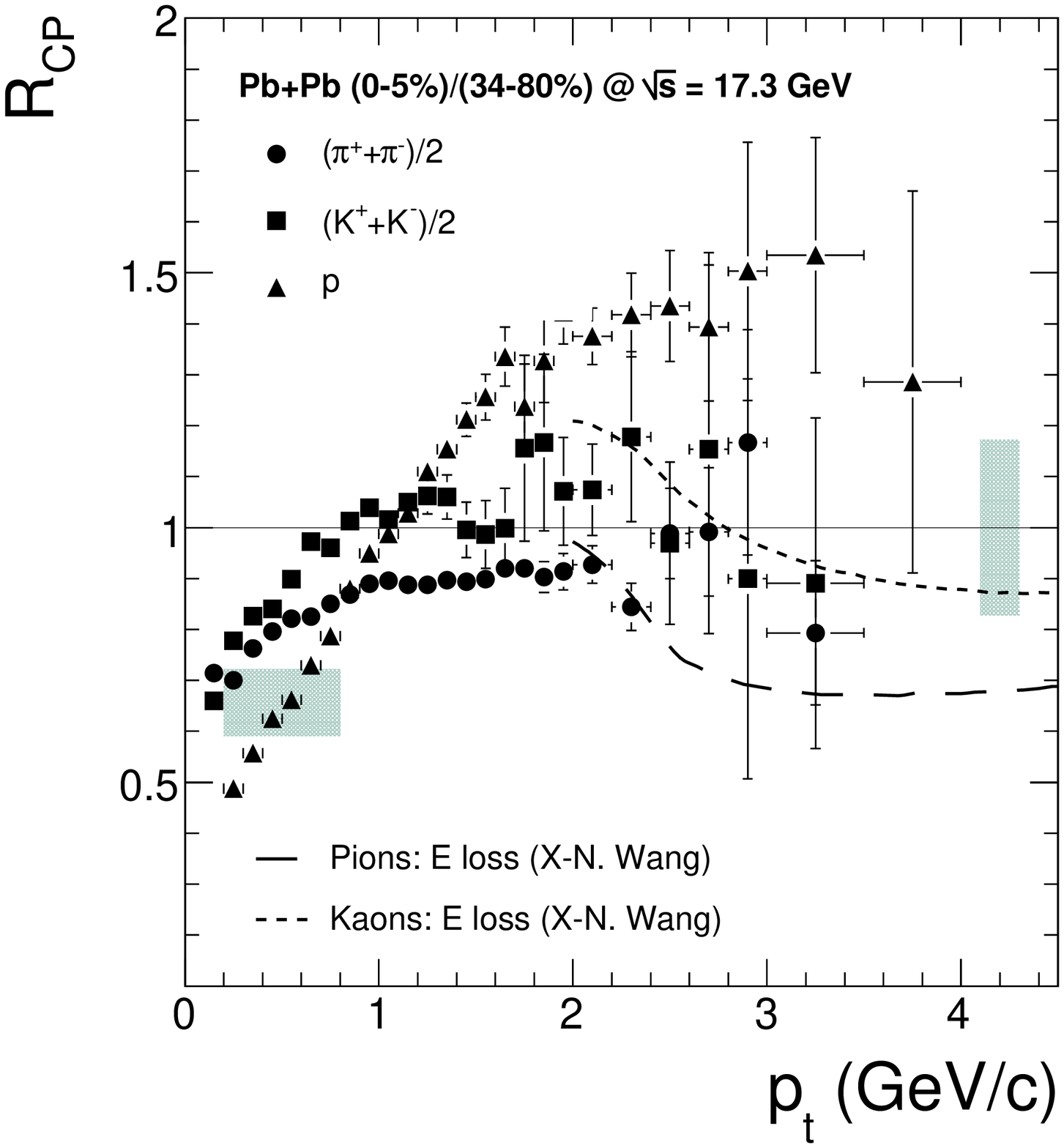}
\end{center}
\end{minipage}
\end{center}
\caption{Left: The nuclear modifcation factor \rcp\ for \kzeros\ as measured by NA57 
\cite{na57a}, compared to predictions with and without parton energy 
loss \cite{wang1,pqm}.
Right. \rcp\ for charged pions, charged kaons, and protons measured by NA49 \cite{na49a}. 
The dashed (dotted) curves are calculations with parton energy loss for pions 
(kaons) \cite{wang1}.}
\label{FIrcp}
\end{figure}

%Figure~\ref{FIraa} displays the nuclear modification factor \raa, defined as:
%\begin{equation}
%  \raa = \frac{\sigma^{\rb{pp}}_{\rb{inel}}}{\ncoll} 
%         \frac{d^{2}N(A + A) / (d\mypt dy)}
%              {d^{2}\sigma(p + p) / (d\mypt dy)}
%\end{equation}
Figure~\ref{FIraa} displays the nuclear modification factor \raa\ for pions
based on central A+A data measured by WA98, NA45, NA49 \cite{wa98pi0,na45c,na49a}.
Here, \raa\ was constructed by using the p+p parametrization of Blattnig et al. 
\cite{blattnig}, as already suggested in \cite{david}. A new feature in this
figure are the recent NA49 data that extend the \mypt-reach up to
4.5~\gevc. Bearing in mind all the caveats described in the previous section, 
the shown \raa\ values are consistent
with the theoretical expectation for an energy loss scenario at realistic gluon 
densities of $dN_{\rb{g}}/dy = 400 - 600$ \cite{glv1}. They are clearly below
the \rpa\ values measured at slighly higher \sqrts, which should provide
a lower bound on the expected \rpa\ at SPS.

\begin{figure}[t]
\begin{center}
\begin{minipage}[b]{68mm}
\begin{center}
\includegraphics[height=78mm]{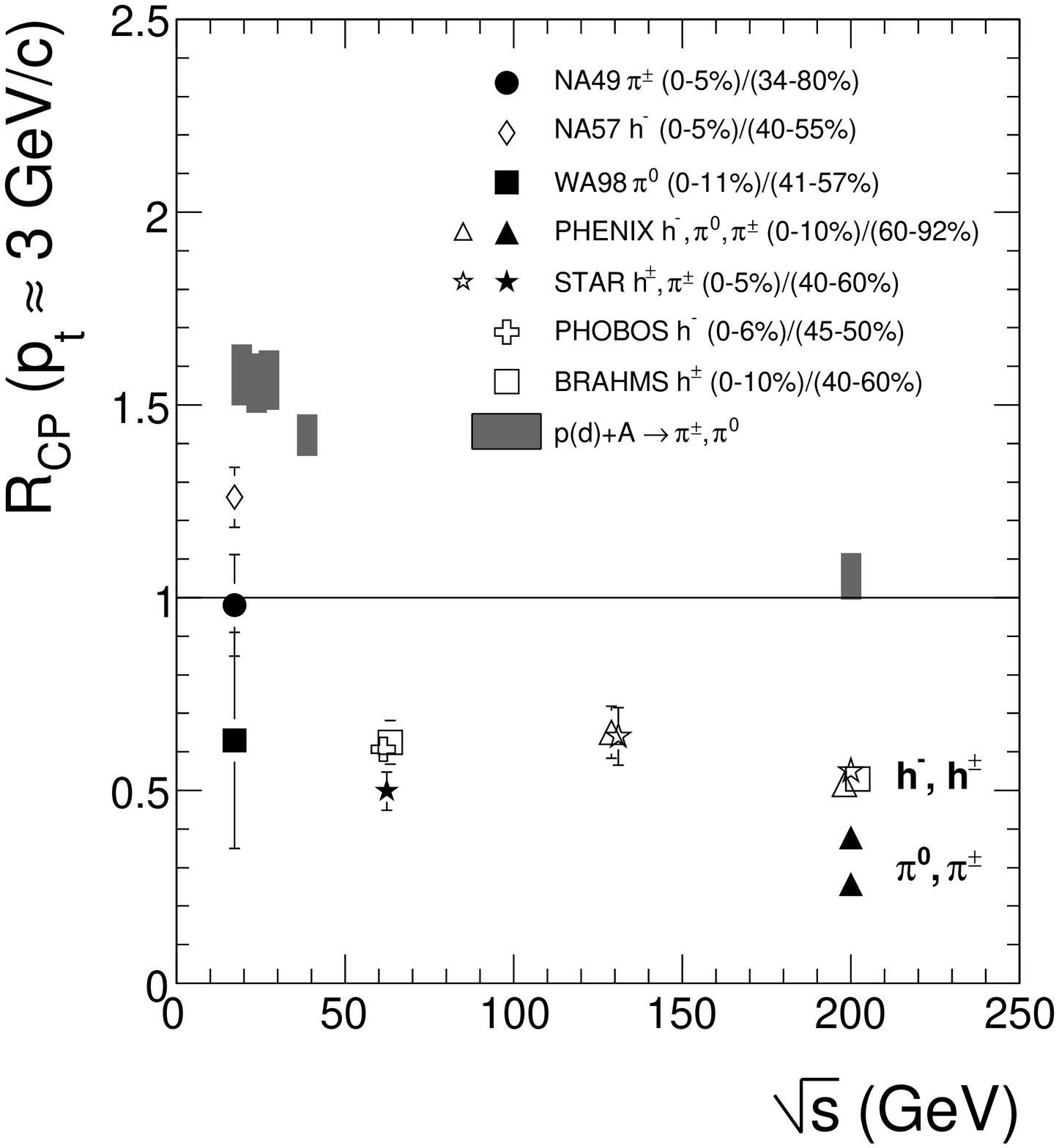}
\end{center}
\end{minipage}
\begin{minipage}[b]{68mm}
\begin{center}
\includegraphics[height=78mm]{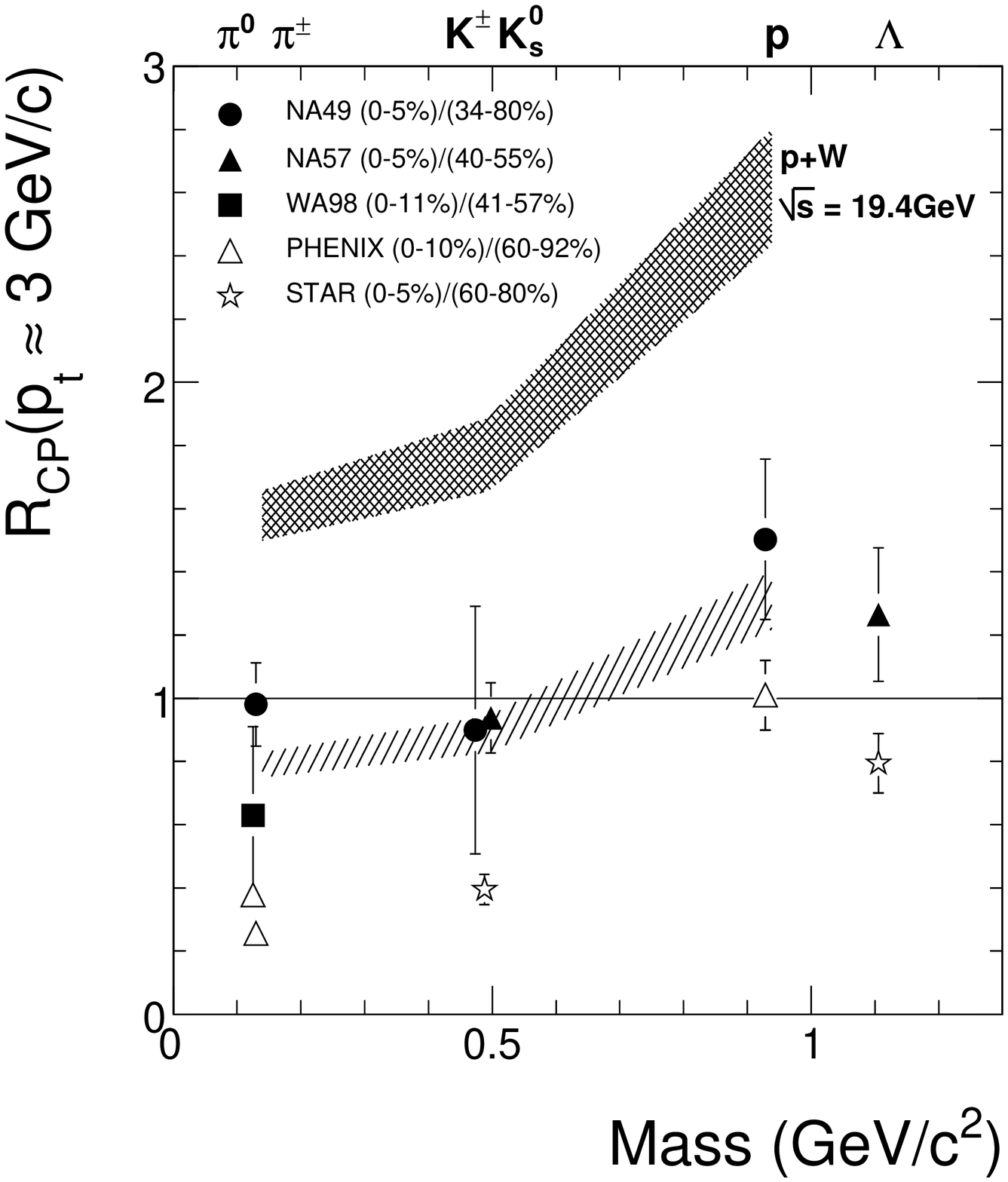}
\end{center}
\end{minipage}
\end{center}
\caption[]{Left: The energy dependence of the nuclear modification factor \rcp\ 
for pions and h$^{\pm}$ at $\mypt \approx 3$~\gevc\ 
\cite{wa98pi0,na49a,na57a,stard,phenixd,stare,starf,phenixe,phenixf,brahmsa,brahmsb,phobosa}. 
Also included is \rpa\ (grey boxes) for p+W at the same \mypt \cite{cronin}.
Right: The dependence of \rcp\ on the particle mass for SPS (\sqrts~=~17.3~GeV, 
filled symbols) \cite{wa98pi0,na49a,na57a} and RHIC (\sqrts~=~200~GeV, 
open symbols) \cite{phenixb,starc}. The dark grey band 
represents \rpa\ from p+W data at \sqrts~=~19.4~GeV \cite{cronin}, while the 
hatched band displays the same values scaled by 0.5.}
\label{FIrcpDep}
\end{figure}

The problem of the missing p+p baseline measurement can to a certain extent be 
circumvented by using peripheral A+A collisions as a reference instead. 
Figure~\ref{FIrcp} shows the \rcp\ for \kzeros\ by NA57 (data is also available for
\hmin, \lam, and \lab \cite{na57a}) and the \rcp\ for pions, kaons, and protons by
NA49 \cite{na49a}. A comparison of the data to calculations including parton
energy loss \cite{wang1,pqm} shows that the measured \rcp\ values are in line with a quenching
scenario. When considering all systematic errors introduced by different 
methods of centrality determination and the estimation of \ncoll, the results of
NA49, NA57, and WA98 are found to be consistent with each other.
The energy dependence of \rcp\ at intermediate \mypt, which is summarized in the 
left panel of \Fi{FIrcpDep}, apparantly is strongest between \sqrts~=~17.3~GeV 
and \sqrts~=~62~GeV. An interesting question is whether this is just a reflection 
of a similar energy dependence of the Cronin effect or due to a dramatic increase 
of the parton energy loss. Data on p(d)+A at \sqrts~=~62~GeV would be needed to 
provide an answer.
As already visible in \Fi{FIrcp}, there is a clear mass dependence of \rcp:
\rcp(p,\lam) $>$ \rcp(K) $\ge$ \rcp($\pi$).
This effect is quite similar at SPS and RHIC (see right panel of \Fi{FIrcpDep}), 
even though the overall scale is naturally lower at higher energies. However, the 
same mass dependence is present in \rpa\ at \sqrts~=~19.4~GeV. Scaled down by a 
factor 0.5 it nicely matches the data for A+A at \sqrts~=~17.3~GeV, while
the corresponding factor is 0.2 at \sqrts~=~200~GeV. 
%This means that the suppression relative to
%p+A at \sqrts~=~17.3~GeV is of the order of 0.5 at $\mypt \approx 3$~\gevc, 
%compared to 0.2 at \sqrts~=~200~GeV. 

\section{High $p_{\rb{t}}$ correlations}

\begin{figure}[ht]
\begin{center}
\includegraphics[width=\textwidth]{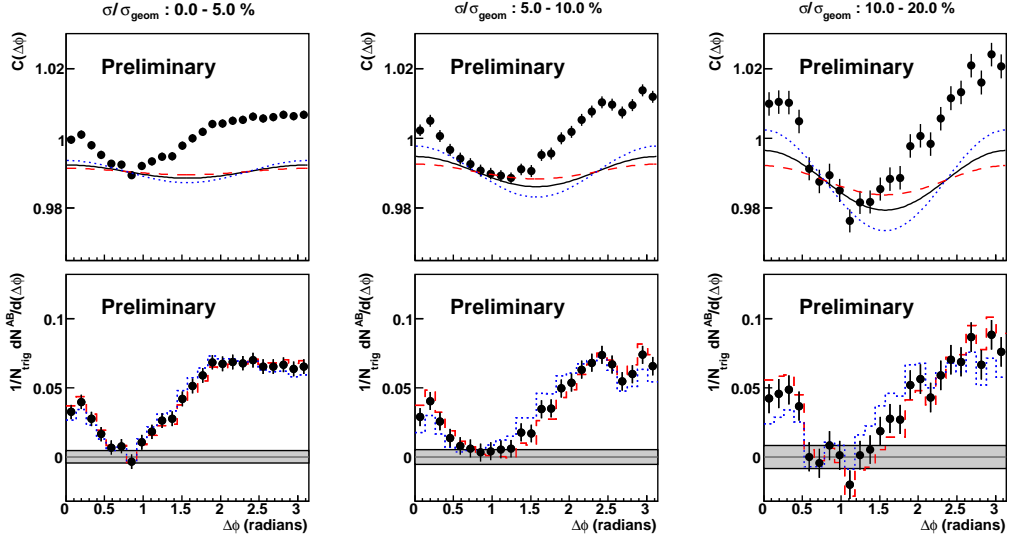}
\end{center}
\caption[]{Upper row: Correlation functions for Pb+Au reactions at \sqrts~=~17.3~GeV
for three centrality bins. The full line represents the flow contribution estimated via 
the ZYAM method \cite{phenixc}. The dashed and dotted lines indicated the statistical 
uncertainty of this contribution.
Lower row: Conditional yields of the jet-pair distribution in three centrality bins,
normalized to the number of triggers \cite{na45b}.}
\label{FIceresCorr}
\end{figure}

Figure~\ref{FIceresCorr} shows the two particle correlation functions $C(\Delta\phi)$, 
as well as the conditional yields of the jet associated hadrons measurend by 
NA45~\cite{na45b} at \sqrts~=~17.3~GeV.
The method and the choice of the \mypt-ranges ($2.5 \:\gevc < p_{\rb{t}}^{trigger} < 4.0 \:\gevc$
and $1.0 \:\gevc < p_{\rb{t}}^{assoc.} < 2.5 \:\gevc$) are identical to an analysis performed
by the PHENIX collaboration at \sqrts~=~200~GeV \cite{phenixc}. 
The away side structure observed at SPS is clearly broader than a p+p like di-jet
expectation, as e.g. predicted by PYTHIA, and the width does not depend significantly on
centrality. However, an indication for a modification of its shape with centrality
is observed: The away-side structure seems to develop a flat top for central events.
Even though this modification of the shape is much stronger at RHIC, this observations
are qualitatively similar at both energies and might in both case be indicative for
partonic interaction with the produced medium.

\section{Baryon-meson ratios at high $p_{\rb{t}}$}

The mass hierarchy observed in the \rcp\ values at SPS might point to the fact
that the quark coalescence approach, that has been proposed to explain the large
baryon/meson ratios at intermediate \mypt\ observed at RHIC \cite{starb}, 
might also be valid at SPS energies. The upper panels of \Fi{FIbmratios} show
the p/\piplus, \pbar/\pimin, and \lam/\kzeros\ ratios as measured by NA49,
together with the corresponding results from STAR \cite{starb} and PHENIX 
\cite{phenixc} at \sqrts~=~200~GeV. Generally, the higher net-baryon
density at SPS energies manifests itself in the differences in the overall
scale of the ratios. What is remarkable though is the fact that shape of the
\mypt-dependence is identical to the one observed at RHIC. The lower panels
of \Fi{FIbmratios} show the double ratios which do not exhibit a significant
\mypt-dependence. A similar observation has been made in the double ratios 
of \rcp\ by the NA57 collaboration \cite{na57b}.
This would indicate that a possible transition from a hydrodynamical 
hadronization picture to quark coalescence would happen in the same 
\mypt-region at SPS and RHIC.

\begin{figure}[t]
\begin{center}
\begin{minipage}[b]{45mm}
\begin{center}
\includegraphics[height=58mm]{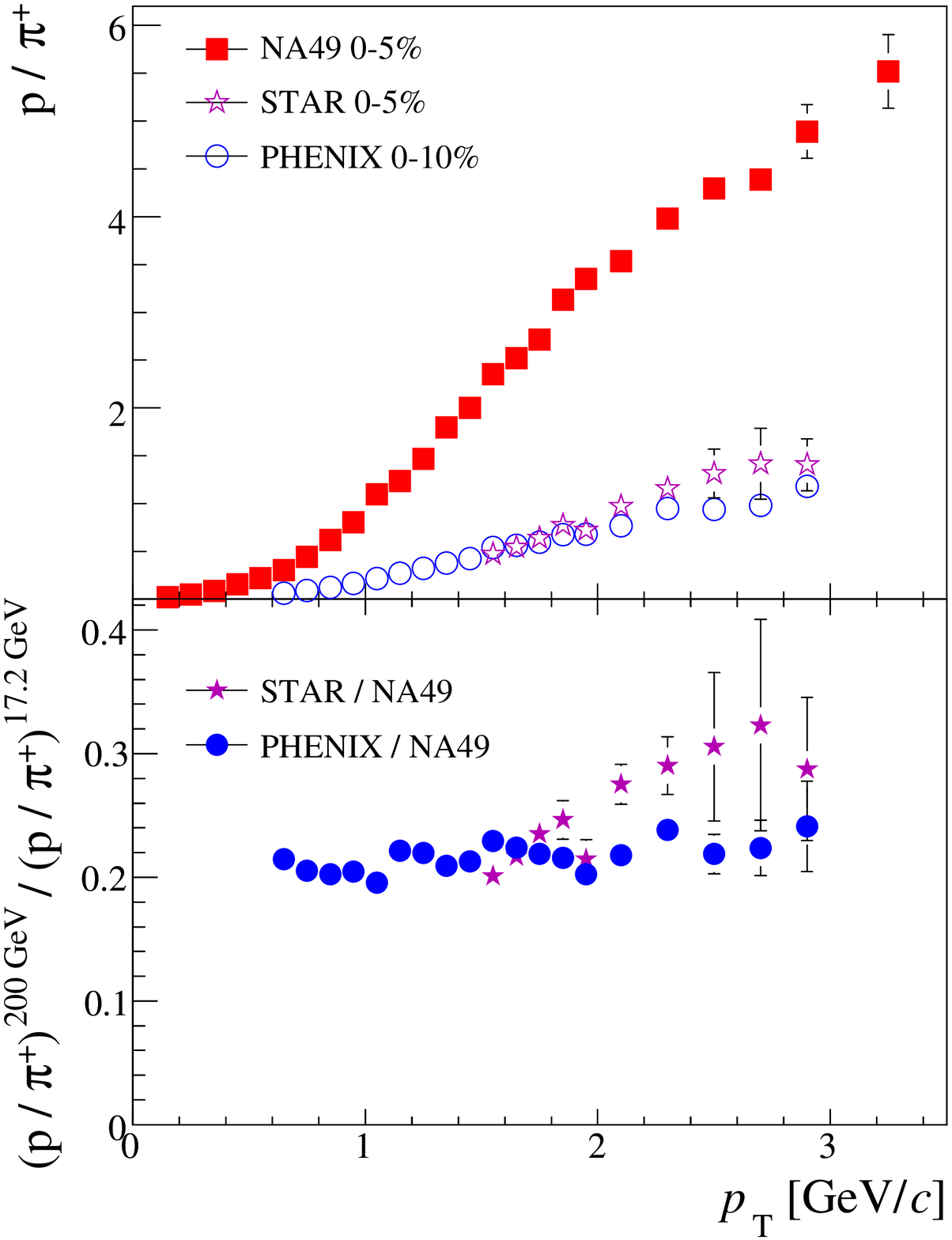}
\end{center}
\end{minipage}
\begin{minipage}[b]{45mm}
\begin{center}
\includegraphics[height=58mm]{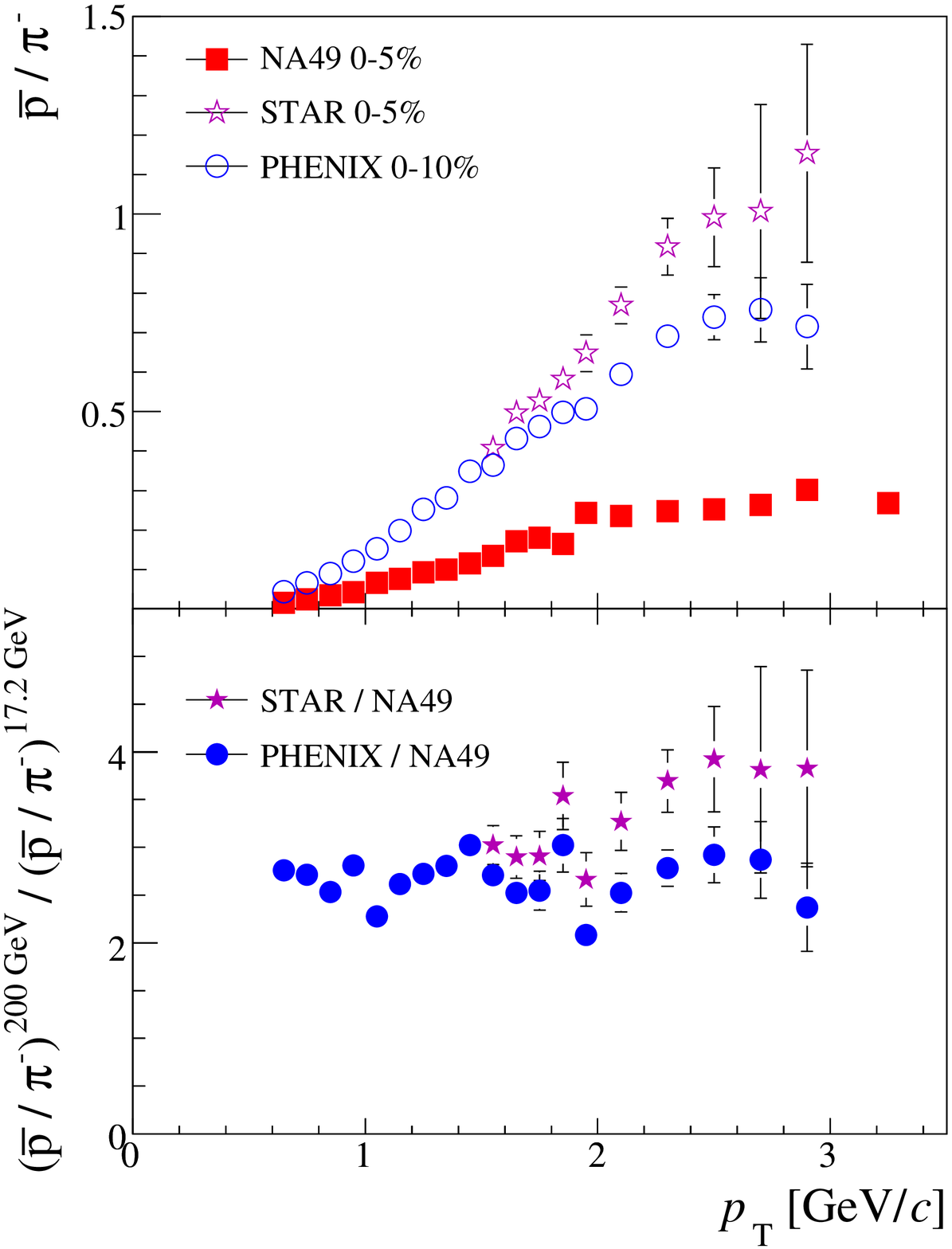}
\end{center}
\end{minipage}
\begin{minipage}[b]{45mm}
\begin{center}
\includegraphics[height=58mm]{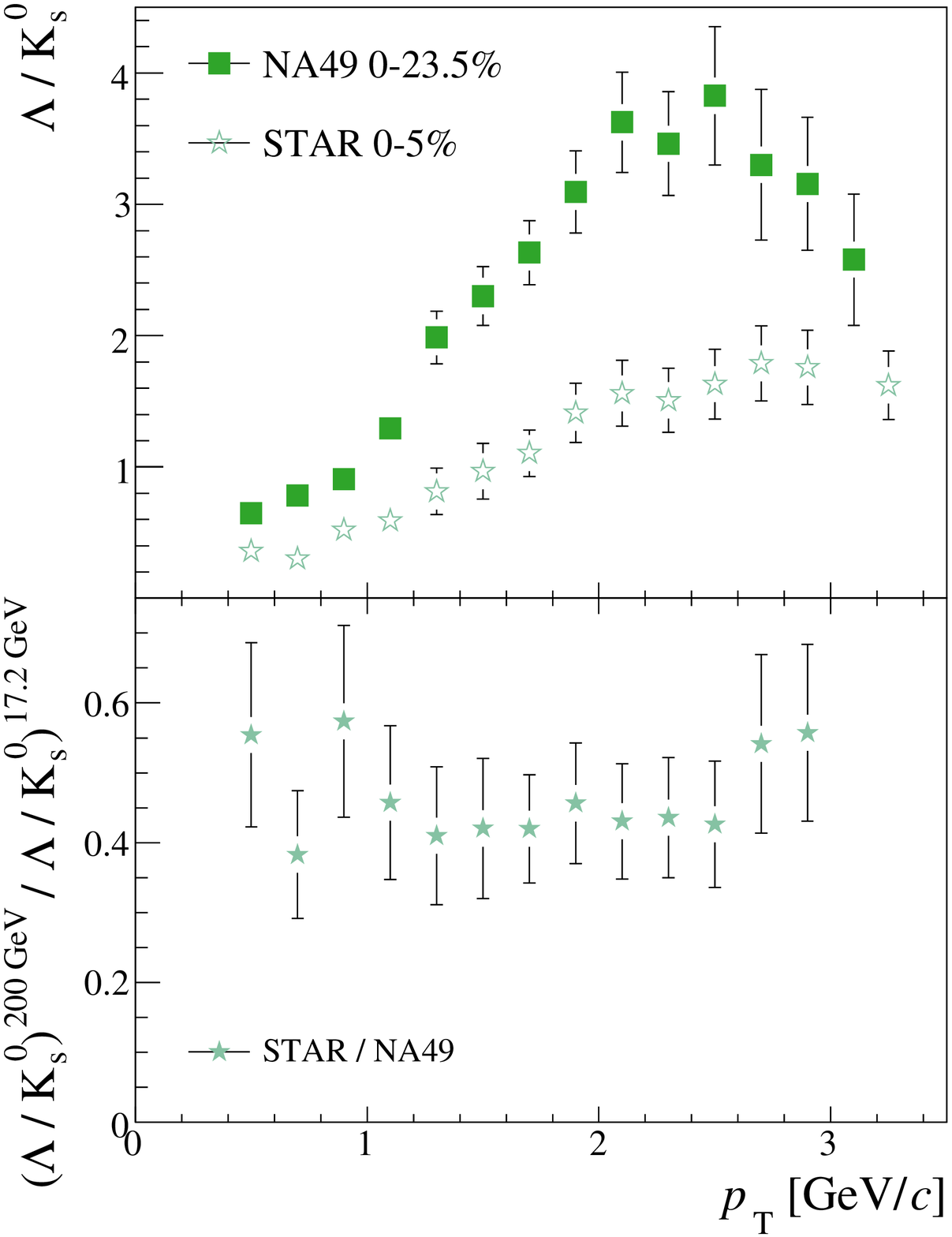}
\end{center}
\end{minipage}
\end{center}
\caption{The baryon/meson ratios as measured by NA49 at \sqrts~=~17.3~GeV \cite{tim},
compared to results from RHIC at \sqrts~=~200~GeV \cite{starb,phenixc}. Please note 
the different scales of the plots.}
\label{FIbmratios}
\end{figure}

\section{Summary}

New data on high \mypt\ hadron production at the CERN SPS were reviewed.
The interpretation of the nucleus--nucleus high \mypt\ data at the SPS generally 
suffers from the lack of reliable p+p and p+A baseline measurements in the 
relevant \mypt-range. This is particularly important here, since the spectral 
shape changes drastically with energy in the SPS region. 
Nevertheless, the general picture that emerges from the new data by NA45, NA49, 
and NA57, as well as from a reassessment of older data (WA98, NA45), is that
also at SPS parton energy loss may be present. The \rcp\ values
show no sign of Cronin enhancement but rather indicate that the Cronin effect
is counterbalanced by jet quenching. 
Also, two-particle azimuthal correlations exhibit a qualitatively similar 
behaviour at SPS than at RHIC and might indicate a partonic interaction with
the medium. The new SPS data on intermediate \mypt\ baryon/meson ratios might help
to understand the underlying hadronization mechanisms.

\appendix
The author would like to thank the organisers of Hard Probes 2006 for the 
invitation to the conference. Also, valuable discussions with D.~d'Enterria
and M.~P{\l}osko\'{n} are greatfully acknowledged.

% \section{}
% \label{}

\end{document}